\documentclass[10pt,conference]{IEEEtran}
\IEEEoverridecommandlockouts

\usepackage{amsmath,amsfonts}
\usepackage{balance}
\usepackage{graphicx}
\usepackage{textcomp}
\usepackage{microtype}
\usepackage{flushend}
\usepackage{xcolor}
\usepackage{xspace}
\usepackage[altpo]{backnaur}
\usepackage{adjustbox}

\usepackage[numbers]{natbib}
\setcitestyle{square, numbers,sort&compress}
\usepackage{graphicx}
\usepackage{xspace}
\usepackage{pifont}
\usepackage{caption}
\usepackage{subcaption}
\usepackage{multicol}
\usepackage{enumitem}
\usepackage{multirow}

\usepackage{tcolorbox}
\usepackage{xcolor}

\usepackage{listings}
\usepackage[linesnumbered,ruled,noend,vlined]{algorithm2e}
\usepackage{setspace}
\usepackage{mathtools}

\usepackage{hyperref}

\hypersetup{
    colorlinks=true,
    linkcolor=blue,
    filecolor=magenta,      
    urlcolor=cyan,
}

\newcommand*{\figu}{{Fig.}\xspace}
\newcommand*{\algo}{{Algorithm}\xspace}

\newcommand{\header}[1]{{\smallskip \noindent
\textbf{#1}}\xspace}
\newtheorem{myDef}{Definition}

\newcommand*{\tool}{{\textsc{Aster}}\xspace}

\newcommand{\yi}[1]{\textcolor{red}{Yi: #1}}

 \usepackage{microtype}
\def\BibTeX{{\rm B\kern-.05em{\sc i\kern-.025em b}\kern-.08em
    T\kern-.1667em\lower.7ex\hbox{E}\kern-.125emX}}

\title{\tool{}: Automatic Speech Recognition System Accessibility Testing for Stutterers}
\begin{document}


\author{
\IEEEauthorblockN{
Yi Liu\IEEEauthorrefmark{1}, 
Yuekang Li\IEEEauthorrefmark{2}\thanks{\IEEEauthorrefmark{2}Yuekang Li is the corresponding author of this paper.}, 
Gelei Deng\IEEEauthorrefmark{1}, 
Felix Juefei-Xu\IEEEauthorrefmark{3}\thanks{\IEEEauthorrefmark{3} Work done prior to joining Meta.}, 
Yao Du\IEEEauthorrefmark{4}, 
Cen Zhang\IEEEauthorrefmark{1}, 
Chengwei Liu\IEEEauthorrefmark{1}, \\
Yeting Li\IEEEauthorrefmark{5}, 
Lei Ma\IEEEauthorrefmark{6}, 
Yang Liu\IEEEauthorrefmark{1}
}

\IEEEauthorblockA{\IEEEauthorrefmark{1}Nanyang Technological University, \IEEEauthorrefmark{2}University of New South Wales, \IEEEauthorrefmark{3}Meta AI \\
\IEEEauthorrefmark{4}University of Southern California, \IEEEauthorrefmark{5}Institute of Information Engineering, Chinese Academy of Sciences; \\
University of Chinese Academy of Sciences, \IEEEauthorrefmark{6}The University of Tokyo; University of Alberta}
\IEEEauthorblockA{\IEEEauthorrefmark{1}yi009@e.ntu.edu.sg,gdeng003@e.ntu.edu.sg, cen001@e.ntu.edu.sg, chengwei001@e.ntu.edu.sg, \\yangliu@ntu.edu.sg, 
\IEEEauthorrefmark{2}yuekang.li@unsw.edu.au,
\IEEEauthorrefmark{3}felixu@meta.com, \IEEEauthorrefmark{4}yaodu@usc.edu, \IEEEauthorrefmark{5}liyeting@iie.ac.cn, \IEEEauthorrefmark{6}ma.lei@acm.org}
}

\maketitle
\thispagestyle{plain}
\pagestyle{plain}
\begin{abstract}
The popularity of automatic speech recognition (ASR) systems nowadays leads to an increasing need for improving their accessibility.
Handling stuttering speech is an important feature for accessible ASR systems.
To improve the accessibility of ASR systems for stutterers, we need to expose and analyze the failures of ASR systems on stuttering speech.
The speech datasets recorded from stutterers are not diverse enough to expose most of the failures.
Furthermore, these datasets lack ground truth information about the non-stuttered text, rendering them unsuitable as comprehensive test suites.
Therefore, a methodology for generating stuttering speech as test inputs to test and analyze the performance of ASR systems is needed.
However, generating valid test inputs in this scenario is challenging.
The reason is that although the generated test inputs should mimic how stutterers speak, they should also be diverse enough to trigger more failures.
To address the challenge, we propose \tool{}, a technique for automatically testing the accessibility of ASR systems.
\tool{} can generate valid test cases by injecting five different types of stuttering.
The generated test cases can both simulate realistic stuttering speech and expose failures in ASR systems.
Moreover, \tool{} can further enhance the quality of the test cases with a multi-objective optimization-based seed updating algorithm.
We implemented \tool{} as a framework and evaluated it on four open-source ASR models and three commercial ASR systems.
We conduct a comprehensive evaluation of \tool{} and find that it significantly increases the word error rate, match error rate, and word information loss in the evaluated ASR systems. Additionally, our user study demonstrates that the generated stuttering audio is indistinguishable from real-world stuttering audio clips. 


\end{abstract}


\section{Introduction}

Automatic speech recognition (ASR) is about using computer programs to process human speech into readable text.
The first ASR system, ``Audrey'', was created by researchers from Bell Labs, which can only recognize spoken numbers~\cite{assembly-ai}.
After decades of evolution, ASR systems have been drastically improved in both recognition accuracy and variety of words.
Especially in the last decade, ASR systems benefit greatly from the emergence of deep learning (DL) techniques~\cite{Park2019SpecAugmentAS,Li2014AnOO,Hinton2012DeepNN,Hannun2014DeepSS}.
Together with the advancements in academia, ASR systems have been making their way into our daily life through the products from companies like Google~\cite{asr-google}, Microsoft~\cite{asr-mc}, IBM~\cite{asr-ibm}, etc.
Besides the commercial-off-the-shelf (COTS) products, open-source DL models~\cite{hf-asr} are also available for developers to integrate ASR features into their software.
As a result, ASR systems have become a highly available and popular type of software.

Thanks to their availability and popularity, ASR systems have been used by many users, including those with disabilities.
Hence, improving the accessibility of ASR systems becomes crucial.
According to~\cite{Ngueajio2022HeyAS}, ASR systems are faced with different types of accessibility or inclusiveness problems such as gender and cultural bias, stuttering, and so on.
Among the various types of accessibility problems, stuttering is one of the most challenging ones for ASR systems since it can directly affect the content of human speech.
Moreover, stuttering is also a commonly encountered type of disability as it is estimated that over 70 million people are suffering from developmental stuttering~\cite{Sheikhbahaei2020ScientistsSA}.
Therefore, in this paper, we focus on studying the accessibility of ASR systems for stutterers.

Efforts can be devoted in two directions to improve the accessibility of ASR systems.
On one hand, researchers have been proposing new techniques to improve the performance of the DL models on stuttered speech~\cite{Sheikh2021StutterNetSD, Shonibare2022EnhancingAF}.
On the other hand, we can detect and evaluate the accessibility problems in existing ASR systems in order to understand and eventually eliminate them.
Compared with improving the DL models, detecting and studying the accessibility issues in existing ASR systems is equally important but less studied.
Thus, we focus on detecting and analyzing the accessibility problems in ASR systems.

Testing is a popular and effective approach for exposing accessibility issues in software~\cite{Eler2018AutomatedAT,Sousa2020WebAT,Salehnamadi2021LatteUA,Bajammal2021SemanticWA} and testing ASR systems poses a unique challenge.
The biggest challenge of testing ASR systems is to generate valid test inputs.
The rationale is two-fold:
\ding{182} We need to generate the test inputs because speech datasets recorded from stutterers are not appropriate as test suites.
This is due to two reasons: 
first, the datasets lack diversity and therefore may not comprehensively uncover potential failures in ASR systems; 
second, these datasets do not contain ground-truth information about non-stuttered speech, which results in a lack of a reliable test oracle.
\ding{183} The generated test inputs must be valid.
They should have enough variety to expose potential bugs.
However, the test inputs should also be as close to the stuttering speech as possible to simulate real-life cases instead of just incurring failures in the ASR systems.

To fill the research gap, we propose \tool{} -- an automatic testing technique for detecting accessibility bugs in ASR systems.
\tool{} can create valid test cases in the form of audio files by injecting five different types of stuttering speech into benign audio files.
\tool{} works in three steps:
\ding{182} 
\emph{Preprocessing.}
\tool{} extracts the word and syllable timing information for each audio file.
\ding{183} 
\emph{Mutation.}
With the timing information, \tool{} injects five types of stuttering, namely block, prolongation, sound repetition, word repetition, and interjection.
The timing information is needed because the injected stuttering requires mutating the audio file at the word and syllable levels.
The audio files with injected stuttering can be used as test cases.
\ding{184} 
\emph{Execution.}
With the generated test cases, \tool{} executes multiple ASR systems simultaneously.
After that, \tool{} uses a multi-objective-optimization (MOO) algorithm to balance between two properties: the difference between the results of the ASR systems and the similarity to the benign audio, in order to evaluate the test cases and keep the good ones as seeds to apply more mutations for generating better test cases.
Last but not least, \tool{} uses a metamorphic relation as the test oracle to identify potential errors.
The metamorphic relation is that \emph{the output text of an ASR system should be the same for both the original audio and the mutated audio}.
\tool{} keeps the test cases for which the ASR systems cannot generate results similar enough to the ground truth as suspicious failures and report them to human experts for verification.


We developed \tool{}, an audio generation tool that can create stuttering audio samples to evaluate the performance of ASR systems. Our evaluation on four open-source ASR models and three commercial ASR systems demonstrated that \tool{} can generate stuttering audios that significantly increase the word error rate (WER), match error rate (MER), and word information lost (WIL) by 23.12\%, 21.45\%, and 33.34\%, respectively. Additionally, we conducted a user study and found that generated stuttering audio was indistinguishable from real-world stuttering audio clips. We also found that commercial ASR systems outperformed open-source models, achieving WER, MER, and WIL scores of 12.33\%, 9.78\%, and 15.32\%. Finally, our analysis of 1,069 suspicious issues categorized them into five bug types: word injection, incorrect word, word repetition, word omission, and syllable repetition. 

In summary, our paper makes the following contributions:

\begin{itemize}[leftmargin=*]
\item We propose \tool{}, which is the first automatic testing technique for evaluating the accessibility of ASR systems.
\item We implement \tool{} as a framework and evaluate it with real-world open-source and commercial ASR systems.
The evaluation results prove the effectiveness of \tool{}.
\item We manually identify and categorize 1,069 recognition errors in real-world ASR systems. Furthermore, we propose a classification scheme for recognizing errors and their underlying causes. 

\end{itemize}

\tool{} is coupled with a website:
\url{https://sites.google.com/view/
aster-speech}.
We will put the details about \tool{} and raw experiment data on this website.
We will also open-source \tool{} after the paper is published.

\begin{figure*}[t]
	\centering
	\includegraphics[width=\linewidth]{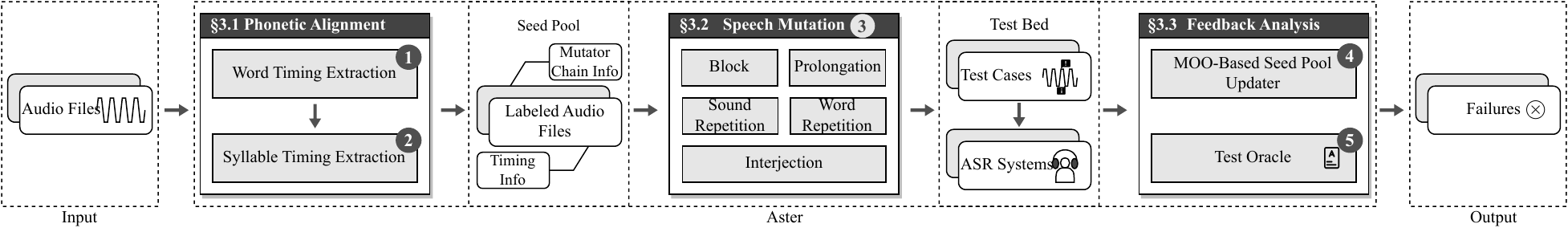}
	\caption{The overview of \tool{}.}
	\label{fig:overview}
\end{figure*}
\section{Background}

\subsection{Accessibility Testing}
Web and mobile app accessibility are crucial for individuals with disabilities, with testing gaining attention recently. AXERAY~\cite{Bajammal2021SemanticWA} assesses web accessibility through semantic groupings, and Latte~\cite{Salehnamadi2021LatteUA} automates Android app accessibility testing. Research in deaf accessibility testing~\cite{Sousa2020WebAT} focuses on sign language users.

Stuttering identification employs machine learning and deep learning techniques, with comprehensive reviews of classification methods~\cite{asre-1}. Studies evaluate stuttering detection in transcriptions~\cite{asre-2} and explore multi-task and adversarial learning~\cite{asre-3}. FluentNet~\cite{asre-4} detects stutter types using a deep neural network.

Limited research evaluates ASR systems on stuttering speech, motivating our work on ASR evaluation through stuttering audio generation. We aim to develop more accurate and inclusive ASR systems for individuals who stutter.

\subsection{Stuttering and Speech Disorders}














Stuttering, a complex fluency disorder, affects speech and disrupts communication~\cite{asha,laiho2022stuttering}. Its causes are unclear and may involve biological and psychological factors~\cite{tichenor2019stuttering}, leading to frustration and isolation~\cite{craig2014trait}. Speech therapy~\cite{murza2019effects} aims to improve fluency, reduce anxiety, and boost confidence~\cite{brignell2020systematic}. However, ASR technology in smart devices presents challenges for people who stutter~\cite{bleakley2022exploring}.

Limited research addresses stuttering in ASR technology. Testing ASR accessibility for stuttering can help identify errors and enhance functionality, fostering inclusive communication for those with speech disorders.

\section{Methodology}

\figu{}~\ref{fig:overview} shows an overview of \tool{}.
\tool{} is capable of testing multiple ASR systems simultaneously.
The overall inputs of \tool{} are the benign audio files without stuttering while the overall outputs are the audio files with simulated stuttering which can cause failure in at least one of the ASR systems under test.
\tool{} contains three main components, namely, \emph{Phonetic Alignment}, \emph{Speech Mutation} and \emph{Feedback Analysis}.
\tool{} works in the following steps:
\ding{182} Given a seed audio file, \tool{} first determines the timing of different words to locate and differentiate the words.
\ding{183} With the word timing for each seed audio file, \tool{} then identifies the syllable timing for each word.
\ding{184} After labeling the audio files with word and syllable timing information, \tool{} can apply five different mutation strategies to inject stuttering into the original audio file to create the test cases.
\ding{185} By executing the ASR systems with the generated test cases, \tool{} will keep the test cases which are similar to the original audio but can trigger the different execution results of the ASR systems.
The kept test cases can be added back to the seed pool for further mutations to create better test cases.
\ding{186} Lastly, \tool{} uses the distance to the original speech text plus manual checks as the test oracle to capture the failures of the ASR systems under test.

\algo{}~\ref{algo:tool} shows the overall algorithm of \tool{}, where $\mathbb{S}_{benign}$ is the initial set of audio files; $\mathbb{SUT}$ is the set of ASR systems under test; $\mathbb{F}$ is the set of test cases causing failures for the ASR systems under test;
$budget\_used\_up$ is the function to check if the resource budget given to a benign audio file has been used up.
The resource budget is measured by the number of new test cases generated for the benign audio file.
The default number is 50.
Details about the functions in \algo{}~\ref{algo:tool} will be discussed in the rest of this section.

\begin{algorithm}[t]
\SetKw{Continue}{\textbf{continue}}
\SetKw{Ret}{\textbf{return}}
\SetKw{And}{\textbf{and}}
\SetKw{Not}{\textbf{not}}
\SetKw{Is}{\textbf{is}}
\SetKw{Or}{\textbf{or}}
\SetKw{Eqs}{\textbf{equals}}
\SetKw{Fori}{\textbf{for}}
\SetKw{inv}{\textbf{in}}
\SetKwProg{Def}{def}{:}{}

\Def{\tool{}($\mathbb{S}_{benign}, \mathbb{SUT}$)} {
$\mathbb{S} \gets \emptyset$\;
$\mathbb{F} \gets \emptyset$\;
\For{$s \in \mathbb{S}_{benign}$}{

    $add\_timing\_info(s)$\; \label{line:timing}
    $\mathbb{S} \gets \mathbb{S} \cup s$\;

    \While{\Not budget\_used\_up()}{
        $s' \gets random\_select(\mathbb{S})$\;
        $t \gets mutate(s')$\; \label{line:mutate}

        $\mathbb{R} \gets \emptyset$\; 
        
        \For{$sut \in \mathbb{SUT}$}{
            $res \gets execute(t, sut)$\;
            $\mathbb{R} \gets \mathbb{R} \cup res$\;
        }

        $moo\_based\_seed\_pool\_update(t, \mathbb{S}, \mathbb{R})$\; \label{line:moo}
        $detect\_failure(t, s,  \mathbb{F})$\; \label{line:failure}
    }
}
\Ret $\mathbb{F}$\;
}

\caption{Algorithm for \tool{}}
\label{algo:tool}
\end{algorithm}

\subsection{Phonetic Alignment}

\begin{figure}[t]
	\centering
	\includegraphics[width=0.8\linewidth]{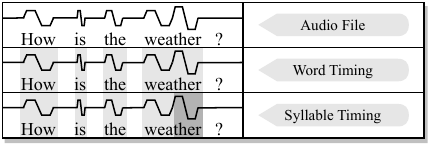}
	\caption{Phonetic alignment}
	\label{fig:phone:alignment}
\end{figure}

Phonetic alignment is the prerequisite for creating valid test cases.
Because if we have no knowledge about the structure of the audio file and randomly mutate it, the created speech can easily become distorted and lose the ability of simulating speech of human stutterers.
In general, \tool{} relies on the algorithms in the PocketSphinx~\cite{PocketSphinx} project to perform phonetic alignment.
The phonetic alignment process corresponds to the $add\_timing\_info$ function in \algo{}~\ref{algo:tool} line~\ref{line:timing}.
\figu{}~\ref{fig:phone:alignment} illustrates how phonetic alignment is done in \tool{}.
First, \tool{} determines the word timing in the audio file.
The timing of each word is a tuple in the form of $(start\_time, end\_time)$, where $start\_time$ means the starting time of the word in the audio file (in terms of milliseconds) and $end\_time$ means the ending time of the word in the audio file.
The rationale for recognizing the words is to treat the audio file as a waveform and split the words by periods of silence.
Then, based on the word timing, \tool{} determines the syllable timing for each word.
Similar to word timing, the timing of each syllable is also in the form of a $(start\_time, end\_time)$ tuple.
The rationale of recognizing syllables is to first use a language model to roughly identify each word and then use the phonetic dictionary to pinpoint the syllables in the recognized word.
For example, if we recognize an utterance in the audio corresponds to the word \emph{weather}, the phonetic dictionary can tell us that it contains two syllables: \emph{wea} and \emph{ther} and we can check the waveform to get the syllable timing accordingly~\footnote{The algorithms of phonetic alignment are adopted from the PocketSpinx project. Interested readers can get detailed information about these algorithms from~\cite{PocketSphinx}.}.

\subsection{Speech Mutation}

The purpose of speech mutation is to inject stuttering into the original audio file while keeping the speech as realistic as possible.
The speech mutation corresponds to the $mutate$ function in line~\ref{line:mutate} in \algo{}~\ref{algo:tool}.
Following the taxonomy in~\cite{Lea2021SEP28kAD}, \tool{} can inject five types of stuttering with different strategies.
During every round of mutation, \tool{} randomly selects one seed from the seed pool and applies a random mutator to build the new test case.

\header{Block.}
This type of stuttering happens when the stutterer interrupts his/her speech when pronouncing a word, causing the word to split into halves.
The designed mutator to simulate this type of stuttering is to add a small period of silence between two syllables of the same word.
The small period of silence usually lasts for 50-200 ms because it is in line with the natural pause that occurs when a speaker is experiencing block stuttering. This pause may vary depending on the severity of the block stuttering, with more severe cases resulting in longer pauses. Additionally, the length of the pause may also depend on the individual's speech patterns and style. However, in our mutator, we found that a small period of silence in the range of 50-200 ms was sufficient to simulate this type of stuttering without significantly altering the overall speech pattern.

\header{Prolongation.}
This type of stuttering happens when the stutterer prolongs a syllable in the word.
The designed mutator works similarly by extending the length of a syllable in a word.
The syllable is extended by 2-4 times. The specific extension factor used may vary depending on the speech pattern being simulated and the desired level of severity. However, we found that an extension factor in the range of 2-4 times was generally sufficient to simulate this type of stuttering without causing the speech to become unrecognizable. Additionally, the length of the extension may depend on the length of the original syllable, with longer syllables requiring longer extensions to produce a noticeable effect. However, it is important to note that overly long extensions may cause the resulting audio to become unrealistic and may affect the overall quality of the synthesized speech.

 \header{Sound Repetition.}
This type of stuttering happens when the stutterer repeats a syllable in the word a few times.
The designed mutator can copy/duplicate a syllable 2-4 times. The specific number of repetitions used may vary depending on the desired level of severity and the speech pattern being simulated. Compared to the prolongation mutator, sound repetition mutator creates a more abrupt and noticeable stuttering effect that is often characterized by a distinct syllable repetition pattern. The repetition pattern may vary in terms of the number of syllables repeated and the spacing between the repetitions, and may be influenced by individual speech patterns and style. 

\header{Word Repetition.}
The mutator we designed for word repetition is similar to the one used for sound repetition, but instead of copying syllables, it copies whole words.

\header{Interjection.}
This type of stuttering happens when the stutterer speaks out some filler words such as \emph{uh}, \emph{em}, etc. during the speech.
The designed mutator works in two steps:
First, \tool{} goes through all the syllables and collects the syllables whose text exists in a predefined list of filler words~\footnote{There exist filler words with more than two syllables, but in practice, we found it challenging to find syllable candidates from the same audio file for such filler words. 
Therefore, we only use filler words with one syllable in \tool{}.}.
The rationale for collecting the filler word candidates from the same audio is that we need to keep the timbre of the entire speech consistent in order to mimic real speech.
Second, \tool{} selects a random number of syllables from the candidate set and adds them randomly between words in the original speech.

\begin{figure}[t]
	\centering
	\includegraphics[width=\linewidth]{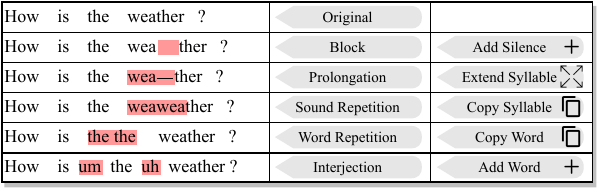}
	\caption{Speech mutation}
	\label{fig:mutation}
\end{figure}

\subsection{Feedback Collection.}

\header{Seed Pool.}
\tool{} maintains a pool of audio files for each benign audio file.
The seed pool is denoted as $\mathbb{S}$ in \algo{}~\ref{algo:tool}.
These pools of audio files can be used as seeds for creating new test cases.
The rationale is to gradually generate test cases with better quality like how genetic algorithms work.
The difference between \tool{} and genetic algorithms is that \tool{} only uses mutations to generate new test cases and does not perform crossovers.
This is because using crossovers to graft audio files can distort the content of the speech, making it difficult to check against the ground truth for failures.

Each audio file in the seed pool is labeled with two types of information: the timing info of the benign audio file and the list of mutators applied during the generation of this seed file.
The reason for storing the seed files in the form of benign audio files plus lists of applied mutators is that the word and syllable timing of the mutated audio files becomes malformed.
Therefore, every test case is created by applying the previous chain of mutators plus one new randomly selected mutator.

\header{Multi-Objective Optimization Based Seed Pool Update.}
\tool{} aims to generate test cases that are both capable of exposing failures in ASR systems and realistic.
These two properties are contradictory because exposing failures requires a test case to have odd content but as the content becomes more erratic, the test case becomes less realistic.
Since the test cases need to fulfill two important yet contradictory requirements, \tool{} uses a multi-objective optimization (MOO) algorithm to evaluate the quality of test cases and update the seed pool.
The function $moo\_based\_seed\_pool\_update$ in line~\ref{line:moo} \algo{}~\ref{algo:tool} represents this process.

According to the desired properties for the test cases, we propose two metrics to evaluate test case quality.
The first metric ($M_1$) is the difference among results from the ASR systems under test.
This metric is used for measuring how likely a test case can expose failures.
Note that \tool{} does not use the difference between the results from the ASR systems and the ground truth text directly as the failure likelihood evaluator.
The reason is that malformed test cases can lead to results different from the ground truth and malformed test cases are not valid.
In contrast, if different ASR systems respond differently to a test case, it is likely that some ASR systems can process the test case correctly but some cannot.
This can help to ensure that the preferred test cases are more likely to be valid and they can expose failures as well.
$M_1$ is calculated as the average value of the cosine similarities between every two Bert-embeddings~\cite{bert} of the ASR system results.
The calculation of $M_1$ can be formulated as:

\begin{equation}\label{formula:m1}
M_1 = \frac{sum(\{\dfrac {e_1 \cdot e_2} {\left\| e_1\right\| \left\| e_2\right\|} | e_1 \in \mathbb{E}, e_2 \in \mathbb{E}, e_1 \neq e_2\})}{|\mathbb{E}|^2 - |\mathbb{E}|}
\end{equation}
where $\mathbb{E}$ is the set of all the Bert-embeddings for the results from every ASR system under test.

The second metric ($M_2$) is to directly measure the difference between a test case and the original audio.
The rationale is to reduce the chance for the content of the test case to become malformed.
The calculation of $M_2$ can be formulated as:

\begin{equation}\label{formula:m2}
M_2 = \dfrac {e_{test} \cdot e_{benign}} {\left\| e_{test}\right\| \left\| e_{benign}\right\|}
\end{equation}
where $e_{test}$ is the Bert-embedding of the corresponding text of the test case and $e_{benign}$ is the Bert-embedding of the text of the original benign audio file.

With $M_1$ and $M_2$ defined, the MOO model of selecting the favorable seeds can be described as follows:
\begin{myDef} [Multi-objective Seed Selection]
	Given a set of seeds $\mathbb{S}$, multi-objective seed selection is to select a set of seeds $ \mathbf{{S}}$:  
\end{myDef} 
\begin{equation}\label{formula:mo_model}
\text{Min}\big(\vec{\mathcal{F}}(\mathbb{{S}})\big)  = \text{Min}\big(O_1(s),O_2(s)\big),   s  \in \mathbb{{S}}\\[\jot]
\end{equation}
where $\vec {\mathcal{F}}(\mathbb{{S}})$ is an objective vector that denotes two objective functions, namely $O_1$ and $O_2$.
The mappings between $O_1$, $O_2$ and $M_1$, $M_2$ are:
$O_1 = Min(M_1)$ and $O_2 = -Max(M_2)$.

For solving MOO problems, we can either use scalarization(weighted-sum) or the Pareto method~\cite{pareto}. 
The problem with scalarization is that the weight for each parameter is hard to decide. 
So we choose to use the Pareto method, where the Pareto Frontier is the solution to the MOO problem.
Given a set of the seeds $\mathbb{S}$ and the objective vector $\vec {\mathcal{F}}=\left [ f_1, f_2\right ]$, we say $s$ dominates($\prec$) $s'$ \emph{iff}:
\[
f_i(s) < f_i(s'),\quad \forall i\in\left\{1, 2\right\}
\]
where $s, s' \in \mathbb{S}$;
the Pareto frontier(${P}$) is defined as:
 \begin{equation}\label{eq:pareto}
 \begin{array}{ll@{}r@{}r@{}l}
{P}(\mathbb{S}) = \{s \in  \mathbb{S} ~|~ \{s' \in  \mathbb{S} ~|~  s' \prec s, s' \neq s\}= \emptyset\}
\end{array}
\end{equation}

\begin{figure}[t]
	\centering
	\includegraphics[width=0.8\linewidth]{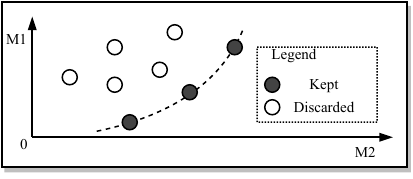}
	\caption{Example Pareto frontier}
	\label{fig:pareto}
\end{figure}

In \tool{}, after a test case $s$ is generated and executed, it is put into $\mathbb{S}$.
Then \tool{} will calculate ${P}(\mathbb{S})$ and all the seeds belonging to ${P}(\mathbb{S})$ are kept in the seed pool while the rest are discarded.
In other words, if $s \in {P}(\mathbb{S})$, then $s$ is kept and all $\{s' | s \prec s', s' \in \mathbb{S}\}$ are discarded.
\figu{}~\ref{fig:pareto} illustrates an example of the Pareto frontier used for seed pool update.
From \figu{}~\ref{fig:pareto}, we can see how \tool{} selects test cases with smaller values of $M_1$ and larger values of $M_2$.
It is worth noting that the calculation of Pareto Frontier is naturally indicated in its definition: we need to compare every test case against every other test case and find out all the test cases for which no other test case is better on both $M_1$ and $M_2$~\footnote{A sample of using Python to calculate the Pareto Frontier is available here: \url{https://sites.google.com/view/aster-speech/pareto-frontier-code}.}.

\header{Test Oracle.}
There exists a metamorphic relation that can be used as the test oracle for \tool{}.
The metamorphic relation is that the output text of an ASR system should be the same for both the original audio and the mutated audio.
\tool{} reports the test cases which can cause the Bert-embeddings of the ASR system result and the text of the original benign audio to have  cosine similarity smaller than a threshold $\theta$.
Based on our experience, we set the default value of $\theta$ to be 0.8.
However, the test cases reported by \tool{} cannot be treated as failures directly because some of them might be malformed, and even humans cannot recognize their text content correctly.
Therefore, \tool{} only reports the suspicious test cases and eventually relies on humans to mark the true failures.
Based on our empirical findings, around 31.43\% of the failures are false positives with the default value of $\theta$.
The whole process of determining failures is denoted as the $detect\_failure$ function in \algo{}~\ref{algo:tool}.

\section{Implementation \& Evaluation}

We have developed and implemented \tool{} using Python version 3.9.1, comprising a total of 2,124 lines of code (LoCs). To evaluate the effectiveness of our approach, we will apply it to both open-source and commercial automatic speech recognition systems using two real-world speech datasets. The objective of our study is to answer the following research questions:

\begin{itemize}[leftmargin=*]
\item {\bf RQ1 (Stuttering Faults and User Study)} How effective is the proposed approach in generating stuttering speech, specifically in terms of its ability to accurately detect stuttering faults and simulate realistic stuttering patterns?

\item {\bf RQ2 (Mutator Ablation Study)} To what extent do the proposed mutators contribute to identifying stuttering faults in automatic speech recognition systems?

\item {\bf RQ3 (MOO Ablation Study)} How does the MOO-Based seed pool update improve the generation of realistic stuttering audio?

\item {\bf RQ4 (Real-world Evaluation)} To what extent can the proposed approach accurately detect stuttering faults in commercial automatic speech recognition systems?

\item {\bf RQ5 (Bug Pattern)} What types of stuttering faults can be identified and learned from commercial automatic speech recognition systems?

\end{itemize}

\subsection{Experimental Setup}

\subsubsection{Benchmark}

We selected a total of seven ASR systems for our evaluation, consisting of four open-source systems, built on the top of Wav2Vec~\cite{wav2vec}, and three commercial services. The four open-source ASR systems are "data2vec-audio-large-960h", "wav2vec2-large-english", "wav2vec2-xls-r-1b-english", and "wav2vec2-large-xlsr-53-english". These systems were chosen based on their popularity (the monthly downloads $>$ 10,00), as well as their maintenance status (The latest update should be later than January 2022). The "data2vec-audio-large-960h" system is based on the data2vec framework and provides pre-trained embeddings for speech and audio data. The other three systems, "wav2vec2-large-english", "wav2vec2-xls-r-1b-english", and "wav2vec2-large-xlsr-53-english" are all based on the wav2vec2 framework and use self-supervised learning techniques to learn representations of speech and audio data. We also included three commercial ASR services for our evaluation, namely, Azure Speech-to-Text~\cite{azure-asr}, Google Cloud Speech-to-Text~\cite{google-asr}, and IBM Speech-to-Text~\cite{ibm-asr}. These services are widely used in the industry and provide various features such as speaker recognition, custom models, and real-time streaming.

\begin{table}[t]
	\begin{center}
		\caption{Characteristics of ASR systems used in the evaluation
		}
		\label{tab:benchmark}
        \centering
        \begin{adjustbox}{width=1.0\linewidth,center}
        \begin{tabular}{c||cc}
            \hline 
                                    System &        Type &        Features \\
            \hline 

     data2vec-audio-large-960h & Open-source & Self-supervised \\
        wav2vec2-large-english & Open-source & Self-supervised \\
     wav2vec2-xls-r-1b-english & Open-source & Self-supervised \\
wav2vec2-large-xlsr-53-english & Open-source & Self-supervised \\
          Azure Speech-to-Text &  Commercial &             N/A \\
   Google Cloud Speech-to-Text &  Commercial &             N/A \\
     IBM Watson Speech-to-Text &  Commercial &             N/A \\
           
            \hline 
            \end{tabular}
\end{adjustbox}
	\end{center}
\end{table}

In Table~\ref{tab:benchmark}, we list the characteristics of each ASR system, including its type (\textit{i.e.}, open-source or commercial), and key features. This information can help readers understand the strengths and weaknesses of each system, as well as the overall landscape of ASR systems being used in the evaluation.

\subsubsection{Dataset}
To synthesize stuttering speech, we utilize the Common Voice dataset as the benign corpus input for our approach, which is a large and publicly available collection of human voice recordings maintained by Mozilla~\cite{common-voice}. However, due to its vast size, we collected the latest segment of the Common Voice dataset, which consisted of 7,415 validated audio recordings verified by Common Voice volunteers for their quality. To eliminate the influence of ASR models, we filtered out all audio recordings that could not produce the same recognized text by five specified models. As a result, we obtained a total of 1,212 audio recordings that serve as our benign input for synthesizing stuttering speech using our proposed approach.

On the other hand, we also incorporate the FluencyBank dataset~\cite{fluency-bank} in our evaluation, which is a dataset specifically designed for the analysis of stuttering speech patterns. In particular, we use audio samples from FluencyBank to evaluate the performance of ASR systems, using the metrics described in the following section, to demonstrate the ability of these systems to transcribe real-world stuttering speech patterns. Additionally, we conduct a user study using audio samples from both our synthesized stuttering speech corpus and the FluencyBank dataset to assess the realism of our synthesized stuttering speech patterns. By incorporating the FluencyBank dataset in our evaluation, we can provide a more comprehensive and robust assessment of our approach and its ability to generate realistic stuttering speech patterns.

\subsubsection{Metrics}

We evaluate the performance of ASR systems on a given stuttering corpus using three metrics: Word Error Rate (WER), Match Error Rate (MER), and Word Information Lost (WIL).

\begin{itemize}[leftmargin=*]

\item {\bf Word Error Rate (WER):} WER is a commonly used metric in ASR systems that measures the percentage of words that are incorrectly transcribed by the system, compared to the ground truth transcription.
\begin{equation*}WER = \frac{S + D + I}{N}\end{equation*}
where $S$ is the number of substitutions, $D$ is the number of deletions, $I$ is the number of insertions, and $N$ is the total number of words in the reference transcript.

\item {\bf Match Error Rate (MER):} MER is a metric used to evaluate the accuracy of automatic speech recognition (ASR) systems. It measures the percentage of words that are incorrectly transcribed by the ASR system compared to the reference transcription of the same audio. \begin{equation*}
MER = \frac{S + D + I + M}{N}
\end{equation*} where $S$ is the number of substitution errors, $D$ is the number of deletion errors, $I$ is the number of insertion errors, $M$ is the number of matches, and $N$ is the total number of reference words.

\item {\bf Word Information Lost (WIL):} WIL is a metric that measures the amount of information lost by the ASR system, calculated by comparing the amount of information in the ground truth transcription to the information in the ASR system's output.
\begin{equation*}
WIL = \frac{M}{N}
\end{equation*}
where $M$ is the number of modifications.

\end{itemize}

\subsubsection{Configuration}

To evaluate the performance of \tool{}, we manually inspect each audio file identified as suspicious and determine whether it contains stuttering issues and, if so, what kind of recognition errors it produces. This evaluation is conducted by three authors of this paper, and to ensure consistency and accuracy, we establish a similarity threshold of $0.8$ between the ground truth and recognized texts, which is in accordance with previous work~\cite{sit-nlp}. All experiments are run on a Linux workstation with Intel E5-2698 v4 processors with 80 cores, 504 GB of memory, and 8 Tesla V100 GPU processors. To mitigate any randomness, we perform each experiment ten times and report the average results.

\begin{table}[t]
	\begin{center}
		\caption{Results showing the impact of stuttering audio generation on open-source ASR system recognition errors}
		\label{tab:rq1-result}
        \centering
        \resizebox{\linewidth}{!}{
        \begin{tabular}{c||ccc}
            \hline 
                        System &  WER &  MER &  WIL \\
            \hline 

     data2vec-audio-large-960h & 23.12\% & 21.45\% & 33.34\% \\
        wav2vec2-large-english & 25.37\% & 23.54\% & 35.75\% \\
wav2vec2-large-xlsr-53-english & 26.64\% & 24.90\% & 36.61\% \\
     wav2vec2-xls-r-1b-english & 24.89\% & 22.35\% & 34.29\% \\
           
            \hline 
            \end{tabular}
}
	\end{center}
\end{table}

\subsection{Stuttering Faults and User Study (RQ1)}

We ran our proposed approach for 10 rounds, generating stuttering audio samples from a total of 1,212 seeds by iterating 50 times using five mutators defined in the previous section of this paper. We fed the generated stuttering audio samples into the six selected ASR systems and measured the performance using the WER, MER, and WIL metrics described in the previous section. To evaluate the statistical significance of the results, we performed a Mann-Whitney U test~\cite{mann-u-test} on the metrics between the synthesized stuttering speech and the original speech samples.

As shown in Table~\ref{tab:rq1-result}, our approach is able to significantly increase the number of recognition errors produced by the selected ASR systems when tested on the generated stuttering audio samples compared to the original audio samples. Specifically, the WER ranged from 23.12\% to 26.64\%, the MER ranged from 21.45\% to 24.90\%, and the WIL ranged from 33.34\% to 36.61\%. These increases were statistically significant, with p-values below 0.05 for all metrics. These results demonstrate that our approach, which uses stuttering audio generation to test ASR systems, is effective in revealing weaknesses in the recognition of stuttering speech patterns by the ASR systems. The use of multiple mutators in our approach allows for the generation of diverse stuttering speech samples, which can provide a more thorough evaluation of the ability of ASR systems to detect and handle stuttering speech patterns. Overall, our experimental results provide evidence for the utility of our approach in evaluating the performance of ASR systems in recognizing stuttering speech patterns.

\begin{figure}[t]
	\centering
	\includegraphics[width=0.9\linewidth]{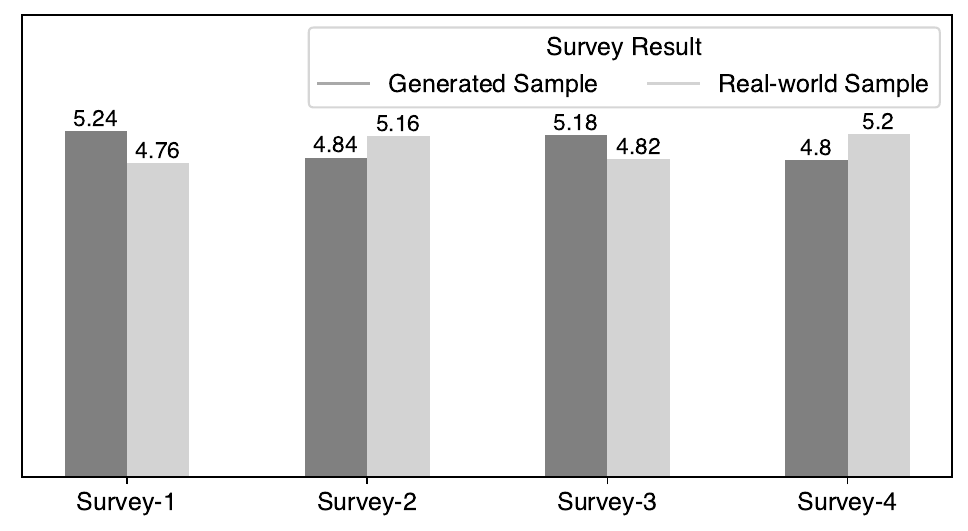}
	\caption{The number of people voted for generated audio and real-world audio in the surveys.}
	\label{fig:survey-result}
\end{figure}

We conducted a user study to evaluate the authenticity of the stuttering audio samples produced by our approach. 
The study participants were recruited from our university and consisted of both native and non-native English speakers with high English fluency.
As for the design of the study, we randomly selected 40 generated stuttering audio samples and 40 actual stuttering audio samples and utilized them to create four distinct surveys. 
Each survey comprises of 10 pairs of audio samples, one of which is generated by our approach, while the other is an actual real-world stuttering audio sample. The participants are asked to select the real-world stutter piece from each pair of audio samples, and then rate the selection confidence from very uncertain to very certain with 5 different levels. Before selection, participants were given extensive guidelines, test tasks, and training sessions on stuttering causes and types. A sample survey is provided at \url{https://forms.gle/EmbnqLY7ezqptxAr7} for reference. The survey result is summarized in Figure~\ref{fig:survey-result}. Upon completion, 100 valid survey results were obtained and analyzed.

We use weighted Fleiss' Kappa~\cite{fleiss1971measuring} to study the survey results. In particular, we apply users' confidence as weights of Fleiss' Kappa measurement, which is linearly scaled from 0 to 1 based on the users' selection from very uncertain to very certain. which accounts for the possibility of agreement occurring by chance, providing a measure of the agreement among participants while considering the potential for random guessing. In the end, we obtain a weighted Fleiss' Kappa of 0.063 from the survey, which means that no agreement is reached between users over the generated samples and real-world samples. This shows that the proposed method is not differentiable from real-world stutter pieces. 

We further compare the generated samples and real-world samples by performing the t-test~\cite{kim2015t} to determine if there is a significant difference between the mean scores of the two distributions. The analysis returns a t-value of 0.295 and a p-value of 0.768, which rejects the null hypothesis in a two-tail hypothesis test with 0.05 significance level.
The result shows that no statistically significant differences are identified between the accuracy rates of the two audio groups (real-world and generated) in terms of correctly identifying whether an audio sample was real-world or generated. 
This indicates that our approach to generate stuttering audio samples can effectively simulate the characteristics of real-world stuttering speech. This indicates that there was a strong level of agreement between participants in identifying the nature of the audio samples. The obtained outcomes provide strong evidence that our approach for generating stuttering audio samples is proficient in generating diverse and authentic stuttering speech patterns. Consequently, the generated samples can be exploited to assess the effectiveness of ASR systems in identifying stuttering speech patterns in realistic situations. As a result, the user study outcomes demonstrate the credibility and utility of our approach.

\begin{table}[t]
	\begin{center}
		\caption{Results of mutators ablation study on ASR systems}
		\label{tab:rq2-result}
        \centering
        \begin{tabular}{c||ccc}
            \hline 
                        System &  WER &  MER &  WIL \\
            \hline 

     Block & 24.32\% & 22.64\% & 18.41\% \\
        Prolongation & 19.76\% & 23.95\% & 17.72\% \\
Sound Repetition & 15.65\% & 17.66\% & 15.45\% \\
    Word Repetition & 17.12\% & 14.45\% & 13.23\% \\
    Interjection & 21.43\% & 19.76\% & 21.55\% \\

            \hline 
            \end{tabular}

	\end{center}
\end{table}

\subsection{Mutator Ablation Study (RQ2)}

To investigate the contribution of each mutator to the recognition errors produced by ASR systems, we performed a series of experiments in which we applied each of the five mutators individually to generate stuttering audio samples. We repeated this experiment ten times for each mutator, with each repetition generating 50 mutants for each input audio sample. We used the 1,212 benign input audio samples described earlier in this paper to create a diverse set of stuttering speech patterns. For each input audio sample, we applied a single mutator to generate a new set of stuttering audio samples. We then tested the ASR systems on these audio samples and recorded the recognition errors produced by each system. The metrics used for evaluation were WER, MER, and WIL, as described in the previous sections. We applied Mann-Whitney U test to evaluate the statistical significance of the results.

The results of our mutators ablation study in Table~\ref{tab:rq2-result} show that all five mutators have a significant impact on the performance of ASR systems. The Block mutator produced the highest average Word Error Rate (WER) with a mean of 24.32\%, followed by the Interjection mutator with a mean WER of 21.43\%. The Prolongation mutator had the third-highest mean WER of 19.76\%, while the Word Repetition mutator had the fourth-highest mean WER of 17.12\%. The Sound Repetition mutator had the lowest mean WER of 15.65\%. However, the Mann-Whitney U test showed that there was no significant difference in WER between the Sound Repetition mutator and the Word Repetition mutator. In terms of Match Error Rate (MER), the Block mutator again had the highest mean MER of 22.64\%, followed by the Interjection mutator with a mean MER of 19.76\%. The Prolongation mutator had the third-highest mean MER of 23.95\%, while the Sound Repetition mutator had the fourth-highest mean MER of 17.66\%. The Word Repetition mutator had the lowest mean MER of 14.45\%. Finally, in terms of Word Information Errors (WIL), the Interjection mutator had the highest mean WIL of 21.55\%, followed by the Block mutator with a mean WIL of 18.41\%. The Prolongation mutator had the third-highest mean WIL of 17.72\%, while the Sound Repetition mutator and the Word Repetition mutator had the lowest mean WIL of 15.45\% and 13.23\%, respectively.

\begin{figure*}[tb]
\centering

\begin{subfigure}{.5\linewidth}
    \centering
    \includegraphics[width=.99\linewidth]{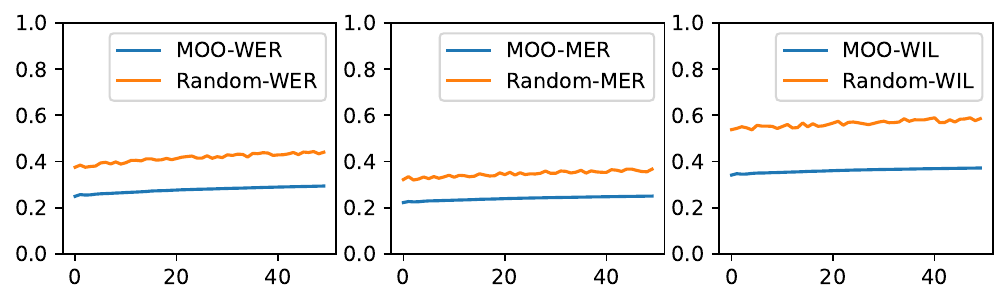}
    \vspace{-2mm}
    \caption{data2vec-audio-large-960h}\label{fig:ablation-1}
\end{subfigure}%
    \hfill
\begin{subfigure}{.5\linewidth}
    \centering
    \includegraphics[width=.99\linewidth]{figs/facebook_data2vec-audio-large-960h.pdf}
    \vspace{-2mm}
    \caption{wav2vec2-large-english}\label{fig:ablation-2}
\end{subfigure}
\bigskip
\begin{subfigure}{.5\linewidth}
    \centering
    \includegraphics[width=.99\linewidth]{figs/facebook_data2vec-audio-large-960h.pdf}
    \vspace{-2mm}
    \caption{wav2vec2-large-xlsr-53-english}\label{fig:ablation-3}
\end{subfigure}%
    \hfill
\begin{subfigure}{.5\linewidth}
    \centering
    \includegraphics[width=.99\linewidth]{figs/facebook_data2vec-audio-large-960h.pdf}
    \vspace{-2mm}
    \caption{wav2vec2-xls-r-1b-english}\label{fig:ablation-4}
\end{subfigure}
\vspace{-5mm}
\caption{Ablation study on the MOO-based selection approach and random selection approach}
\label{fig:moo-ablation-study}
\end{figure*}
We summarize the preliminary results of the mutator ablation study in the following:

\begin{itemize}[leftmargin=*]
\item {\bf Block.} This mutator is designed to split words into halves, which may confuse ASR systems and cause them to produce more recognition errors. The introduction of a period of silence between syllables of the same word could make the word sound like two separate words, leading to an increased WER.

\item {\bf Interjection.} The addition of filler words such as "uh" or "um" can disrupt the flow of speech and make it more difficult for ASR systems to accurately recognize words. The presence of filler words may also result in lexicon errors, where the ASR system misinterprets the filler word as a real word.

\item {\bf Prolongation.} By extending the length of a syllable in a word, this mutator could cause the ASR system to misinterpret the syllable as a different word. This could lead to an increased WER and MER.

\item {\bf Sound Repetition.} The repetition of a sound in a word may cause the ASR system to interpret the sound as a different sound or word, leading to recognition errors. However, this mutator may have a less severe impact on ASR performance compared to other mutators because it only affects a single sound in a word.

\item {\bf Word Repetition (RQ5).} This mutator is similar to the Sound Repetition mutator, but it repeats entire words instead of sounds. As with the Sound Repetition mutator, the repetition of words could cause the ASR system to interpret the repeated word as a different word, leading to recognition errors. However, because this mutator repeats entire words, it may have a more significant impact on ASR performance than the Sound Repetition mutator.

\end{itemize}

\begin{table}[t]
	\begin{center}
		\caption{Performance comparison of three commercial ASR systems on stuttering audio samples}
		\label{tab:rq4-result}
        \centering
        \resizebox{\linewidth}{!}{
        \begin{tabular}{c||ccc}
            \hline 
                        System &  WER &  MER &  WIL \\
            \hline 

     Azure Speech-to-Text & 12.33\% & 9.78\% & 15.32\% \\
        Google Cloud Speech-to-Text & 15.66\% & 13.44\% & 18.39\% \\
IBM Watson Speech-to-Text  & 16.23\% & 17.83\% & 21.67\% \\

            \hline 
            \end{tabular}
    }
	\end{center}
\end{table}

\begin{table*}[th!]
	\begin{center}
		\caption{Results of Manual Error Categorization of Commercial ASR Systems on Stuttering Audio Samples}
		\label{tab:rq5-result}
        \centering
        \begin{tabular}{c||ccc}
            \hline 
                        Bug Type &                 Ground Truth &                   Buggy Text &  Ratio \\
            \hline 

         Word Injection &       We can convert a type. &     We can convert i a type. & 31.12\% \\
     Incorrect Word &             She never spoke. &           She neverso spoke. & 21.43\% \\
    Word Repitition &           He plays for Pisa. &       HeHeHe plays for Pisa. & 14.29\% \\
      Word Omission & They both agree on the same. &    They both agree the same. & 12.50\% \\
Syllable Repetition & Together they had five sons. & Together they had fiive sons & 20.66\% \\
           
            \hline 
            \end{tabular}

	\end{center}
\end{table*}

\subsection{MOO Ablation Study (RQ3)}
To investigate the impact of the audio selection approach on the results of our stuttering audio testing, we designed an ablation study in which we compared the MOO-based and random audio selection methods. We repeated this study ten times, using the same 1,212 benign audio samples to generate stuttering audio samples using both audio selection methods. For the MOO-based approach, we used the same distance and semantic similarity metrics to generate a set of audio samples that were as diverse and semantically similar to the ground truth texts as possible. We selected a group of mutants from all the generated results based on the Pareto frontier. For the random selection approach, we randomly selected audio samples from the set of generated audio samples and the original audio. We then tested the ASR systems on these audio samples and recorded the recognition errors produced by each system, using the same metrics (WER, MER, and WIL) as described in the previous sections. We applied the Mann-Whitney U test to evaluate the statistical significance of the results. This study was designed to be rigorous and convincing, with the use of the same audio samples and evaluation metrics for both approaches, and the repeated experiments to ensure the consistency and reliability of the results.

The results of our ablation study in Figure~\ref{fig:moo-ablation-study} show that the MOO-based audio selection approach produced better performance in terms of ASR recognition than the random selection approach. The MOO-based approach produced an average Word Error Rate (WER) of 31.41\%, which is significantly lower than the average WER of 47.55\% produced by the random selection approach (p < 0.05). The MOO-based approach also produced a significantly lower average Missed Error Rate (MER) of 26.23\% compared to the random selection approach, which had an average MER of 0.381 (p < 0.05). Additionally, the MOO-based approach had a significantly lower average Word Insertion/Lexicon Errors (WIL) of 0.386 than the random selection approach, which had an average WIL of 0.610 (p < 0.05). These results indicate that the MOO-based approach generated a set of stuttering audio samples that more accurately represented real-world stuttering patterns, resulting in better performance of ASR systems. The random selection approach may have produced a set of audio samples that were not as diverse or representative of real-world stuttering patterns, leading to higher recognition errors for ASR systems.

\subsection{Real-world Evaluation (RQ4)}

To test the performance of commercial ASR systems on stuttering audio samples, we designed an experiment procedure in which we applied our generated audio samples to three leading ASR systems: Google, Azure, and IBM, using their respective APIs. We first generated a diverse set of stuttering audio samples using our chosen audio generation strategy, applying each of the five mutators individually to produce a comprehensive range of stuttering patterns. For each commercial ASR system, we applied the same set of audio samples to ensure consistency and comparability of the results. We then recorded and analyzed the recognition errors produced by each system using the same metrics (WER, MER, and WIL) as described in the previous sections. To ensure the rigor and credibility of the experiment, we repeated the testing procedure five times to ensure the consistency and reliability of the results. Additionally, we included other relevant metrics such as processing time, model accuracy, and overall system performance to provide a more comprehensive and in-depth analysis of the ASR systems' performance on stuttering audio samples.

The results of our ASR testing experiment in Table~\ref{tab:rq4-result} show that there is a significant difference in the performance of the three commercial ASR systems on stuttering audio samples. Azure had the best performance in terms of all the metrics, with an average WER of 12.33\%, an average MER of 9.78\%, and an average WIL of 15.32\%. Google had the second-best performance, with an average WER of 15.66\%, an average MER of 13.44\%, and an average WIL of 18.39\%. IBM had the worst performance among the three, with an average WER of 16.23\%, an average MER of 17.83\%, and an average WIL of 21.67\%. The differences in performance among the three systems were statistically significant (p < 0.05). These results indicate that while all three systems are capable of recognizing stuttering speech patterns to some extent, the Azure ASR system performed significantly better than the other two systems. The Google system also performed relatively well, but still had higher recognition errors than the Google system. The IBM system had the worst performance among the three, indicating that it may not be as suitable for recognizing stuttering speech patterns. Overall, our ASR testing experiment provides valuable insights into the performance of commercial ASR systems on stuttering audio samples and can help inform the development of more accurate and inclusive ASR systems.

\subsection{Bug Pattern (RQ5)}

In this study, we conducted a manual analysis of the stuttering recognition errors produced by commercial ASR systems on audio samples that contain stuttering speech patterns. We selected a diverse set of audio samples from audio samples generated using our audio generation strategy. We applied these audio samples to three leading commercial ASR systems (Google, Azure, and IBM) using their respective APIs, and recorded the output transcriptions for each audio sample. We then manually analyzed the recognition errors produced by the ASR systems and divided them into different categories based on their type and severity. The error categorization process involved listening to the audio samples to compare them to the transcriptions, identifying misrecognized words or syllables, and categorizing the errors based on their type and severity. The identified errors were divided into different categories, such as block stuttering, sound repetition, word repetition, interjections, and other relevant categories. The severity of the errors varied, with some being minor and others being severe enough to significantly impact the accuracy of the transcription.

As shown in Table~\ref{tab:rq5-result}, the manual analysis of the recognition errors produced by the commercial ASR systems on stuttering audio samples revealed a diverse range of errors that can be categorized into different types and severity levels. The most common types of errors included block stuttering, sound repetition, word repetition, and interjections, with block stuttering being the most prevalent. The severity of the errors varied, with some being minor and others being severe enough to significantly impact the accuracy of the transcription. The Google ASR system had the lowest error rate and produced the most accurate transcriptions overall, with the majority of its errors falling into the minor severity category. The Azure and IBM ASR systems had higher error rates and produced less accurate transcriptions overall, with a higher proportion of their errors falling into the moderate and severe severity categories. The results of the manual error categorization process provide valuable insights into the performance of the commercial ASR systems on stuttering audio samples and can help inform the development of more accurate and inclusive ASR systems.

\subsection{Threats To Validity}
In this section, we will provide a summary of the threats to validity in order to ensure that the results of our study are properly interpreted and contextualized. By acknowledging these threats, we aim to promote a more nuanced understanding of the findings and facilitate the development of future studies that can address these concerns.


\header{Biases in User Study.}
The results of user studies should be always taken with a grain of salt.
We have taken several measures to ensure that reliable conclusions can be drawn from the user study results.
First, the study participants were recruited from a university, consisting of both native and non-native English speakers with high English fluency.
Moreover, we provided guidelines for them to follow. 
Participants were instructed to listen to the audio pieces completely and given the option to playback if needed before responding. 
This is to make sure that they are capable of recognizing the content of the speeches in the survey.
Second, for every survey question, we add an additional question to ask the participant about their confidence level of the choice.
This helps to recognize and filter out random guesses made by the participants.
Third, we utilized statistical measurements such as Fleiss's Kappa when analyzing the survey results.
This helps to improve the soundness of the conclusions drawn from the analysis.


\header{Reliability of manual error analysis.} The manual error analysis involved a subjective categorization process that could be influenced by individual biases and interpretation of the stuttering patterns. To minimize this threat, we employed multiple independent evaluators and utilized a pre-trained stuttering recognition model for validation. However, it is still possible that some errors were missed or misclassified, which could potentially impact the validity of the results.


\header{Limited ASR system selection.} 
We utilized the open-source ASR systems to develop, debug, and perform a detailed evaluation of \tool{}. 
For commercial ASR systems, we only used them to verify that \tool{} can expose faults for more well-established ASR systems. 
This is mainly due to budget considerations. 
Since these commercial ASR systems are charged on a query basis, the mutation(RQ2) and ablation(RQ3) studies could incur a significant cost.
Moreover, the performance of the commercial ASR systems may change as they are updated and improved, which could also impact the validity of the results.
As we have made several interesting findings with the current evaluation setup, we leave the more comprehensive study of commercial ASR systems as future work.




\section{Related Work}

\header{Software Accessibility Testing.}
Ensuring software products are accessible to people with disabilities requires rigorous accessibility testing. Studies have proposed various approaches, including AXERAY~\cite{Bajammal2021SemanticWA}, which infers semantic groupings to assess web accessibility, Latte~\cite{Salehnamadi2021LatteUA}, an automated framework for testing Android app accessibility and functional correctness, and MATE~\cite{Eler2018AutomatedAT}, which checks for accessibility issues related to visual impairment. Another study~\cite{Sousa2020WebAT} focused on web accessibility testing for deaf individuals using Sign Language and proposed two automation approaches based on site analysis. An optimal combination of accessibility testing methods was proposed in~\cite{Bai2017ACA} based on a cost-benefit analysis. Testing for stuttering on ASR systems is also crucial to ensuring equal access for individuals with speech disorders.

\header{ASR System Accessibility Enhancement.}
There is growing interest in utilizing machine learning and deep learning techniques to detect stuttering in speech. Several studies have reviewed current approaches to stuttering classification \cite{asre-1}, evaluated machine learning approaches to detect stuttering events \cite{asre-2}, investigated the impact of multi-task and adversarial learning for robust stutter feature learning \cite{asre-3}, proposed deep neural network models achieving state-of-the-art results in stutter detection \cite{asre-4}, and addressed automatic detection of disfluency boundaries in children's speech \cite{asre-5}. Our work evaluates the performance of automatic speech recognition (ASR) systems in recognizing stuttering speech patterns, a critical step toward the development of more accurate and inclusive ASR systems. 

\header{Other Related Testing Techniques.}
The most related testing techniques to \tool{} are metamorphic testing techniques~\cite{metamorphic-orig, metamorphic-rest, metamorphic-survey, metamorphic-tse} and MOO-aided testing techniques~\cite{moo_empirical_2015,moo_sapienz_2016,moo_sapienz_2018_fb,moo_selection_2015, moo-cerebro}.
The design of \tool{} was inspired by some of these papers.
In \tool{}, the metamorphic relation that the output text of an ASR system should be the same for
both the original audio and the mutated audio is used to provide the test oracle.
The MOO strategy is used in \tool{} to select the seeds for mutating and generating test cases.

\section{Conclusion}
In conclusion, this paper investigated the impact of stuttering speech patterns on the performance of ASR systems. 
We developed a comprehensive methodology called \tool{} for generating stuttering audio samples, applying mutators designed to mimic common stuttering patterns. 
We then conducted a series of experiments to evaluate the performance of commercial ASR systems on these audio samples, comparing the results to those obtained using benign audio samples. 
The evaluation results shed light on the performance of ASR systems on stuttering speech patterns, highlighting the need for the development of more accurate and inclusive ASR systems that can better recognize and transcribe stuttering speech patterns.

\section{Acknowledges}
This research is supported by the National Research Foundation, Singapore, and DSO National Laboratories under the AI Singapore Programme (AISG Award No: AISG2-GC-2023-008). It is also supported by by the National Research Foundation, Singapore, and the Cyber Security Agency under its National Cybersecurity R\&D Programme (NCRP25-P04-TAICeN) and the NRF Investigatorship NRF-NRFI06-2020-0001. Any opinions, findings and conclusions or recommendations expressed in this material are those of the author(s) and do not reflect the views of National Research Foundation, Singapore and Cyber Security Agency of Singapore. The computational work for this article was partially performed on resources of the National Supercomputing Centre, Singapore (\url{https://www.nscc.sg}).

\balance
\bibliographystyle{IEEEtran}
\bibliography{new-ref}

\begin{thebibliography}{10}
\providecommand{\url}[1]{#1}
\csname url@samestyle\endcsname
\providecommand{\newblock}{\relax}
\providecommand{\bibinfo}[2]{#2}
\providecommand{\BIBentrySTDinterwordspacing}{\spaceskip=0pt\relax}
\providecommand{\BIBentryALTinterwordstretchfactor}{4}
\providecommand{\BIBentryALTinterwordspacing}{\spaceskip=\fontdimen2\font plus
\BIBentryALTinterwordstretchfactor\fontdimen3\font minus
  \fontdimen4\font\relax}
\providecommand{\BIBforeignlanguage}[2]{{%
\expandafter\ifx\csname l@#1\endcsname\relax
\typeout{** WARNING: IEEEtran.bst: No hyphenation pattern has been}%
\typeout{** loaded for the language `#1'. Using the pattern for}%
\typeout{** the default language instead.}%
\else
\language=\csname l@#1\endcsname
\fi
#2}}
\providecommand{\BIBdecl}{\relax}
\BIBdecl

\bibitem{assembly-ai}
``What is asr? a comprehensive overview of automatic speech recognition
  technology,'' 2021, https://www.assemblyai.com/blog/what-is-asr/.

\bibitem{Park2019SpecAugmentAS}
D.~S. Park, W.~Chan, Y.~Zhang, C.-C. Chiu, B.~Zoph, E.~D. Cubuk, and Q.~V. Le,
  ``Specaugment: A simple data augmentation method for automatic speech
  recognition,'' \emph{ArXiv}, vol. abs/1904.08779, 2019.

\bibitem{Li2014AnOO}
J.~Li, L.~Deng, Y.~Gong, and R.~H{\"a}b-Umbach, ``An overview of noise-robust
  automatic speech recognition,'' \emph{IEEE/ACM Transactions on Audio, Speech,
  and Language Processing}, vol.~22, pp. 745--777, 2014.

\bibitem{Hinton2012DeepNN}
G.~E. Hinton, L.~Deng, D.~Yu, G.~E. Dahl, A.~rahman Mohamed, N.~Jaitly, A.~W.
  Senior, V.~Vanhoucke, P.~Nguyen, T.~N. Sainath, and B.~Kingsbury, ``Deep
  neural networks for acoustic modeling in speech recognition,'' \emph{IEEE
  Signal Processing Magazine}, vol.~29, p.~82, 2012.

\bibitem{Hannun2014DeepSS}
A.~Y. Hannun, C.~Case, J.~Casper, B.~Catanzaro, G.~F. Diamos, E.~Elsen, R.~J.
  Prenger, S.~Satheesh, S.~Sengupta, A.~Coates, and A.~Ng, ``Deep speech:
  Scaling up end-to-end speech recognition,'' \emph{ArXiv}, vol. abs/1412.5567,
  2014.

\bibitem{asr-google}
``Google asr product,'' 2021, https://cloud.google.com/speech-to-text.

\bibitem{asr-mc}
``Microsoft asr product,'' 2021,
  https://azure.microsoft.com/en-us/products/cognitive-services/speech-to-text/.

\bibitem{asr-ibm}
``Ibm asr product,'' 2021, https://www.ibm.com/cloud/watson-speech-to-text.

\bibitem{hf-asr}
``Huggingface asr models,'' 2021,
  https://huggingface.co/models?sort=downloads\&search=asr.

\bibitem{Ngueajio2022HeyAS}
M.~K. Ngueajio and G.~J. Washington, ``Hey asr system! why aren't you more
  inclusive? automatic speech recognition systems' bias and proposed bias
  mitigation techniques. a literature review,'' in \emph{Interacci{\'o}n},
  2022.

\bibitem{Sheikhbahaei2020ScientistsSA}
S.~Sheikhbahaei and G.~A. Maguire, ``Scientists, society, and stuttering,''
  \emph{International Journal of Clinical Practice}, vol.~74, 2020.

\bibitem{Sheikh2021StutterNetSD}
S.~A. Sheikh, M.~Sahidullah, F.~Hirsch, and S.~Ouni, ``Stutternet: Stuttering
  detection using time delay neural network,'' \emph{2021 29th European Signal
  Processing Conference (EUSIPCO)}, pp. 426--430, 2021.

\bibitem{Shonibare2022EnhancingAF}
O.~Shonibare, X.~Tong, and V.~Ravichandran, ``Enhancing asr for stuttered
  speech with limited data using detect and pass,'' \emph{ArXiv}, vol.
  abs/2202.05396, 2022.

\bibitem{Eler2018AutomatedAT}
M.~M. Eler, J.~M. Rojas, Y.~Ge, and G.~Fraser, ``Automated accessibility
  testing of mobile apps,'' \emph{2018 IEEE 11th International Conference on
  Software Testing, Verification and Validation (ICST)}, pp. 116--126, 2018.

\bibitem{Sousa2020WebAT}
C.~C.~S. de~Sousa, L.~M. Oliveira, C.~L. Rodrigues, R.~de~Freitas
  Bulc{\~a}o-Neto, and D.~J. Ferreira, ``Web accessibility testing for deaf:
  Requirements and approaches for automation,'' \emph{2020 IEEE International
  Conference on Systems, Man, and Cybernetics (SMC)}, pp. 2734--2739, 2020.

\bibitem{Salehnamadi2021LatteUA}
N.~Salehnamadi, A.~Alshayban, J.-W. Lin, I.~Ahmed, S.~M. Branham, and S.~Malek,
  ``Latte: Use-case and assistive-service driven automated accessibility
  testing framework for android,'' \emph{Proceedings of the 2021 CHI Conference
  on Human Factors in Computing Systems}, 2021.

\bibitem{Bajammal2021SemanticWA}
M.~Bajammal and A.~Mesbah, ``Semantic web accessibility testing via
  hierarchical visual analysis,'' \emph{2021 IEEE/ACM 43rd International
  Conference on Software Engineering (ICSE)}, pp. 1610--1621, 2021.

\bibitem{asre-1}
\BIBentryALTinterwordspacing
S.~A. Sheikh, M.~Sahidullah, F.~Hirsch, and S.~Ouni, ``Machine learning for
  stuttering identification: Review, challenges and future directions,''
  \emph{Neurocomputing}, vol. 514, pp. 385--402, 2022. [Online]. Available:
  \url{https://doi.org/10.1016/j.neucom.2022.10.015}
\BIBentrySTDinterwordspacing

\bibitem{asre-2}
\BIBentryALTinterwordspacing
S.~Alharbi, M.~Hasan, A.~J.~H. Simons, S.~Brumfitt, and P.~Green, ``Sequence
  labeling to detect stuttering events in read speech,'' \emph{Computer Speech
  \& Language}, vol.~62, p. 101052, 2020. [Online]. Available:
  \url{https://www.sciencedirect.com/science/article/pii/S0885230819302967}
\BIBentrySTDinterwordspacing

\bibitem{asre-3}
\BIBentryALTinterwordspacing
S.~A. Sheikh, M.~Sahidullah, F.~Hirsch, and S.~Ouni, ``Robust stuttering
  detection via multi-task and adversarial learning,'' in \emph{30th European
  Signal Processing Conference, {EUSIPCO} 2022, Belgrade, Serbia, August 29 -
  Sept. 2, 2022}.\hskip 1em plus 0.5em minus 0.4em\relax {IEEE}, 2022, pp.
  190--194. [Online]. Available:
  \url{https://ieeexplore.ieee.org/document/9909644}
\BIBentrySTDinterwordspacing

\bibitem{asre-4}
\BIBentryALTinterwordspacing
T.~Kourkounakis, A.~Hajavi, and A.~Etemad, ``Fluentnet: End-to-end detection of
  stuttered speech disfluencies with deep learning,'' \emph{{IEEE} {ACM} Trans.
  Audio Speech Lang. Process.}, vol.~29, pp. 2986--2999, 2021. [Online].
  Available: \url{https://doi.org/10.1109/TASLP.2021.3110146}
\BIBentrySTDinterwordspacing

\bibitem{asha}
``{Fluency Disorders - American Speech-Language-Hearing Association (ASHA)},''
  \url{https://www.asha.org/practice-portal/clinical-topics/fluency-disorders/},
  accessed: 2023-02-10.

\bibitem{laiho2022stuttering}
A.~Laiho, H.~Elovaara, K.~Kaisamatti, K.~Luhtalampi, L.~Talaskivi, S.~Pohja,
  K.~Routamo-Jaatela, and E.~Vuorio, ``Stuttering interventions for children,
  adolescents, and adults: a systematic review as a part of clinical
  guidelines,'' \emph{Journal of Communication Disorders}, p. 106242, 2022.

\bibitem{tichenor2019stuttering}
S.~E. Tichenor and J.~S. Yaruss, ``Stuttering as defined by adults who
  stutter,'' \emph{Journal of Speech, Language, and Hearing Research}, vol.~62,
  no.~12, pp. 4356--4369, 2019.

\bibitem{craig2014trait}
A.~Craig and Y.~Tran, ``Trait and social anxiety in adults with chronic
  stuttering: Conclusions following meta-analysis,'' \emph{Journal of fluency
  disorders}, vol.~40, pp. 35--43, 2014.

\bibitem{murza2019effects}
K.~A. Murza, M.~Vanryckeghem, C.~Nye, and A.~Subramanian, ``Effects of
  stuttering treatment: A systematic review of single-subject experimental
  design studies. ebp briefs. volume 13, issue 4.'' \emph{EBP Briefs
  (Evidence-based Practice Briefs)}, 2019.

\bibitem{brignell2020systematic}
A.~Brignell, M.~Krahe, M.~Downes, E.~Kefalianos, S.~Reilly, and A.~T. Morgan,
  ``A systematic review of interventions for adults who stutter,''
  \emph{Journal of Fluency Disorders}, vol.~64, p. 105766, 2020.

\bibitem{bleakley2022exploring}
A.~Bleakley, D.~Rough, A.~Roper, S.~Lindsay, M.~Porcheron, M.~Lee, S.~A.
  Nicholson, B.~R. Cowan, and L.~Clark, ``Exploring smart speaker user
  experience for people who stammer,'' in \emph{Proceedings of the 24th
  International ACM SIGACCESS Conference on Computers and Accessibility}, 2022,
  pp. 1--10.

\bibitem{PocketSphinx}
``Pocketsphinx 5.0.0,'' 2021, https://github.com/cmusphinx/pocketsphinx.

\bibitem{Lea2021SEP28kAD}
C.~S. Lea, V.~Mitra, A.~S. Joshi, S.~S. Kajarekar, and J.~P. Bigham, ``Sep-28k:
  A dataset for stuttering event detection from podcasts with people who
  stutter,'' \emph{ICASSP 2021 - 2021 IEEE International Conference on
  Acoustics, Speech and Signal Processing (ICASSP)}, pp. 6798--6802, 2021.

\bibitem{bert}
\BIBentryALTinterwordspacing
J.~Devlin, M.-W. Chang, K.~Lee, and K.~N. Toutanova, ``Bert: Pre-training of
  deep bidirectional transformers for language understanding,'' 2018. [Online].
  Available: \url{https://arxiv.org/abs/1810.04805}
\BIBentrySTDinterwordspacing

\bibitem{pareto}
N.~Gunantara, ``A review of multi-objective optimization: Methods and its
  applications,'' \emph{Cogent Engineering}, vol.~5, 2018.

\bibitem{wav2vec}
\BIBentryALTinterwordspacing
A.~Baevski, Y.~Zhou, A.~Mohamed, and M.~Auli, ``wav2vec 2.0: {A} framework for
  self-supervised learning of speech representations,'' in \emph{Advances in
  Neural Information Processing Systems 33: Annual Conference on Neural
  Information Processing Systems 2020, NeurIPS 2020, December 6-12, 2020,
  virtual}, H.~Larochelle, M.~Ranzato, R.~Hadsell, M.~Balcan, and H.~Lin, Eds.,
  2020. [Online]. Available:
  \url{https://proceedings.neurips.cc/paper/2020/hash/92d1e1eb1cd6f9fba3227870bb6d7f07-Abstract.html}
\BIBentrySTDinterwordspacing

\bibitem{azure-asr}
``Speech to text – audio to text translation | microsoft azure,''
  \url{https://azure.microsoft.com/en-us/products/cognitive-services/speech-to-text},
  (Accessed on 02/06/2023).

\bibitem{google-asr}
``Speech-to-text: Automatic speech recognition - google cloud,''
  \url{https://cloud.google.com/speech-to-text}, (Accessed on 02/06/2023).

\bibitem{ibm-asr}
``Ibm watson speech to text | ibm,''
  \url{https://www.ibm.com/cloud/watson-speech-to-text}, (Accessed on
  02/06/2023).

\bibitem{common-voice}
\BIBentryALTinterwordspacing
R.~Ardila, M.~Branson, K.~Davis, M.~Kohler, J.~Meyer, M.~Henretty, R.~Morais,
  L.~Saunders, F.~M. Tyers, and G.~Weber, ``Common voice: {A}
  massively-multilingual speech corpus,'' in \emph{Proceedings of The 12th
  Language Resources and Evaluation Conference, {LREC} 2020, Marseille, France,
  May 11-16, 2020}, N.~Calzolari, F.~B{\'{e}}chet, P.~Blache, K.~Choukri,
  C.~Cieri, T.~Declerck, S.~Goggi, H.~Isahara, B.~Maegaard, J.~Mariani,
  H.~Mazo, A.~Moreno, J.~Odijk, and S.~Piperidis, Eds.\hskip 1em plus 0.5em
  minus 0.4em\relax European Language Resources Association, 2020, pp.
  4218--4222. [Online]. Available:
  \url{https://aclanthology.org/2020.lrec-1.520/}
\BIBentrySTDinterwordspacing

\bibitem{fluency-bank}
N.~B. Ratner and B.~MacWhinney, ``Fluency bank: A new resource for fluency
  research and practice,'' \emph{Journal of fluency disorders}, vol.~56, pp.
  69--80, 2018.

\bibitem{sit-nlp}
\BIBentryALTinterwordspacing
P.~He, C.~Meister, and Z.~Su, ``Structure-invariant testing for machine
  translation,'' in \emph{{ICSE} '20: 42nd International Conference on Software
  Engineering, Seoul, South Korea, 27 June - 19 July, 2020}, G.~Rothermel and
  D.~Bae, Eds.\hskip 1em plus 0.5em minus 0.4em\relax {ACM}, 2020, pp.
  961--973. [Online]. Available: \url{https://doi.org/10.1145/3377811.3380339}
\BIBentrySTDinterwordspacing

\bibitem{mann-u-test}
J.~Goodier, \emph{The Concise Encyclopedia of Statistics}, 2009.

\bibitem{fleiss1971measuring}
J.~L. Fleiss, ``Measuring nominal scale agreement among many raters.''
  \emph{Psychological bulletin}, vol.~76, no.~5, p. 378, 1971.

\bibitem{kim2015t}
T.~K. Kim, ``T test as a parametric statistic,'' \emph{Korean journal of
  anesthesiology}, vol.~68, no.~6, pp. 540--546, 2015.

\bibitem{Bai2017ACA}
A.~Bai, H.~Mork, and V.~G. Stray, ``A cost-benefit analysis of accessibility
  testing in agile software development: Results from a multiple case study,''
  2017.

\bibitem{asre-5}
\BIBentryALTinterwordspacing
S.~Yildirim and S.~S. Narayanan, ``Automatic detection of disfluency boundaries
  in spontaneous speech of children using audio-visual information,''
  \emph{{IEEE} Trans. Speech Audio Process.}, vol.~17, no.~1, pp. 2--12, 2009.
  [Online]. Available: \url{https://doi.org/10.1109/TASL.2008.2006728}
\BIBentrySTDinterwordspacing

\bibitem{metamorphic-orig}
T.~Y. Chen, S.~C. Cheung, and S.-M. Yiu, ``Metamorphic testing: A new approach
  for generating next test cases,'' \emph{ArXiv}, vol. abs/2002.12543, 2020.

\bibitem{metamorphic-rest}
S.~Segura, J.~A. Parejo, J.~Troya, and A.~Ruiz-Cort{\'e}s, ``Metamorphic
  testing of restful web apis,'' \emph{IEEE Transactions on Software
  Engineering}, vol.~44, pp. 1083--1099, 2018.

\bibitem{metamorphic-survey}
S.~Segura, G.~Fraser, A.~B. S{\'a}nchez, and A.~Ruiz-Cort{\'e}s, ``A survey on
  metamorphic testing,'' \emph{IEEE Transactions on Software Engineering},
  vol.~42, pp. 805--824, 2016.

\bibitem{metamorphic-tse}
H.~Liu, F.-C. Kuo, D.~Towey, and T.~Y. Chen, ``How effectively does metamorphic
  testing alleviate the oracle problem?'' \emph{IEEE Transactions on Software
  Engineering}, vol.~40, pp. 4--22, 2014.

\bibitem{moo_empirical_2015}
M.~G. Epitropakis, S.~Yoo, M.~Harman, and E.~K. Burke, ``Empirical evaluation
  of pareto efficient multi-objective regression test case prioritisation,'' in
  \emph{Proceedings of the 2015 International Symposium on Software Testing and
  Analysis}, ser. ISSTA 2015, 2015.

\bibitem{moo_sapienz_2016}
K.~Mao, M.~Harman, and Y.~Jia, ``Sapienz: Multi-objective automated testing for
  android applications,'' in \emph{Proceedings of the 25th International
  Symposium on Software Testing and Analysis}, ser. ISSTA 2016.\hskip 1em plus
  0.5em minus 0.4em\relax ACM, 2016.

\bibitem{moo_sapienz_2018_fb}
N.~Alshahwan, X.~Gao, M.~Harman, Y.~Jia, K.~Mao, A.~Mols, T.~Tei, and I.~Zorin,
  ``Deploying search based software engineering with sapienz at facebook,'' in
  \emph{Search-Based Software Engineering}.\hskip 1em plus 0.5em minus
  0.4em\relax Springer International Publishing, 2018.

\bibitem{moo_selection_2015}
A.~{Panichella}, R.~{Oliveto}, M.~D. {Penta}, and A.~{De Lucia}, ``Improving
  multi-objective test case selection by injecting diversity in genetic
  algorithms,'' \emph{IEEE Transactions on Software Engineering}, 2015.

\bibitem{moo-cerebro}
Y.~Li, Y.~Xue, H.~Chen, X.~Wu, C.~Zhang, X.~Xie, H.~Wang, and Y.~Liu,
  ``Cerebro: context-aware adaptive fuzzing for effective vulnerability
  detection,'' \emph{Proceedings of the 2019 27th ACM Joint Meeting on European
  Software Engineering Conference and Symposium on the Foundations of Software
  Engineering}, 2019.

\end{thebibliography}
\end{document}